\documentclass[journal=nalefd,manuscript=letter]{achemso}

\usepackage[utf8]{inputenc}
\usepackage[T1]{fontenc} 
\usepackage[english]{babel}

\usepackage{amsmath}
\usepackage{amssymb}
\usepackage{graphicx}
\usepackage{hyperref}

\makeatletter
\addto\extrasenglish{%
}
\makeatother

\author{J.-D. Pillet}
\affiliation[Collège de France]
{$\Phi_{0}$, JEIP, USR 3573 CNRS, Collège de France, PSL University, 11, place Marcelin Berthelot, 75231 Paris Cedex 05, France}
\alsoaffiliation[École Polytechnique]
{LSI, CEA/DRF/IRAMIS, Ecole polytechnique, CNRS, Institut Polytechnique de Paris, F-91128 Palaiseau, France}
\author{V. Benzoni}
\author{J. Griesmar}
\author{J.-L. Smirr}
\author{Ç. Ö. Girit}
\email{caglar.girit@college-de-france.fr}
\affiliation[Collège de France]
{$\Phi_{0}$, JEIP, USR 3573 CNRS, Collège de France, PSL University, 11, place Marcelin Berthelot, 75231 Paris Cedex 05, France}

\title{Nonlocal Josephson effect in Andreev molecules}


\begin{document}

\begin{tocentry}

\includegraphics[width=\textwidth]{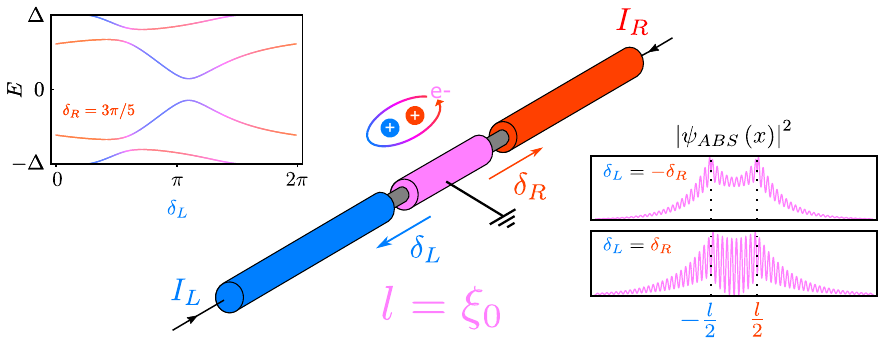}




\end{tocentry}


\begin{abstract}
  
  We propose the ``Andreev molecule,'' an artificial quantum system composed of two closely spaced Josephson junctions.
  The coupling between Josephson junctions in an Andreev molecule occurs through the overlap and hybridization of the junction's ``atomic'' orbitals, Andreev Bound States.
  A striking consequence is that the supercurrent flowing through one junction depends on the superconducting phase difference across the other junction.
  Using the Bogolubiov-de-Gennes formalism, we derive the energy spectrum and non-local current-phase relation for arbitrary separation.
  We demonstrate the possibility of creating a $\varphi$-junction and propose experiments to verify our predictions.
  Andreev molecules may have potential applications in quantum information, metrology, sensing, and molecular simulation.
\end{abstract}


Understanding and exploiting the interaction between Josephson junctions is paramount for superconducting device applications in quantum information~\cite{wendin_2017}, magnetometry~\cite{clarke_2006}, metrology~\cite{benz_2004}, and quantum simulation~\cite{solano_2018}.
In typical superconducting circuits, junctions interact indirectly via electromagnetic coupling to inductors, capacitors, transmission lines, and microwave resonators.
In addition to this well understood long-range interaction~\cite{likharev_dynamics_1986}, there is a short range interaction via quasiparticle diffusion which can modify superconducting energy gaps and critical currents, but is only important close to $T_{c}$, the superconducting transition temperature, or at large bias voltages~\cite{hansen_static_1984}.

A second short-range interaction, mediated by Cooper pairs, is relevant to the majority of applications where characteristic energies are much smaller than the gap, but is still poorly understood.
It becomes significant when the distance between Josephson junctions is comparable to $\xi_0$, the superconducting coherence length, and can modify the electrical properties in a dramatic way.

Initially, minor effects resulting from this “order-parameter interaction'' were calculated for temperatures near $T_{c}$ using the Ginzburg-Landau equations~\cite{deminova_calculation_1979}.
More recently, theorists have investigated this problem at arbitrary temperature using Green's function techniques.
In the two-electrode geometry, where it is not possible to independently apply a phase difference to each junction, the overall current-phase relation and dc current were obtained~\cite{brinkman_2000, golubov_2004}.
For the more relevant three-electrode geometry, non-local
out-of-equilibrium supercurrents were calculated and the existence of $\pi$ shifts in the current-phase relation were demonstrated~\cite{freyn_production_2011,jonckheere_multipair_2013,melin_d.c._2014_2,feinberg_quartets_2015,melin_gate-tunable_2016,melin_simple_2017}.
A remarkable phase-locking similar to Shapiro steps was predicted and subsequently measured experimentally in superconducting bi-junctions biased with commensurate voltages\cite{pfeffer_subgap_2014,cohen_nonlocal_2018}.
The authors attribute these phenomena to the formation of entangled Cooper pairs called ``quartets.''

Although this interpretation is intriguing, many open questions remain: what is the microscopic mechanism involved?
How does the interaction depend on the distance between the junctions?
Have all non-local effects been revealed?
How can one detect these effects and exploit them in devices?

In this work we answer these questions and show that the phenomena can be explained by a simple analogy to the formation of a hydrogen molecule, complete with orbital hybridization and level splitting.
The analog of the hydrogen atom's atomic orbitals are Andreev Bound States (ABS), localized electronic states that are the microscopic basis for the Josephson effect~\cite{bretheau_exciting_2013,janvier_coherent_2015}.
The equivalent of the Bohr radius is $\xi_0$, the characteristic length scale for the spread of ABS wavefunctions.
In our superconducting ``Andreev molecule,'' where two Josephson junctions are separated by a distance $l \lesssim \xi_0$, the ABS wavefunctions localized at each junction overlap and hybridize.
This hybridization modifies the total energy spectrum and can result in a $\varphi$-junction where the supercurrent flowing through one junction depends on the superconducting phase difference across the other junction.

\begin{figure}
  \includegraphics[width=0.9\textwidth]{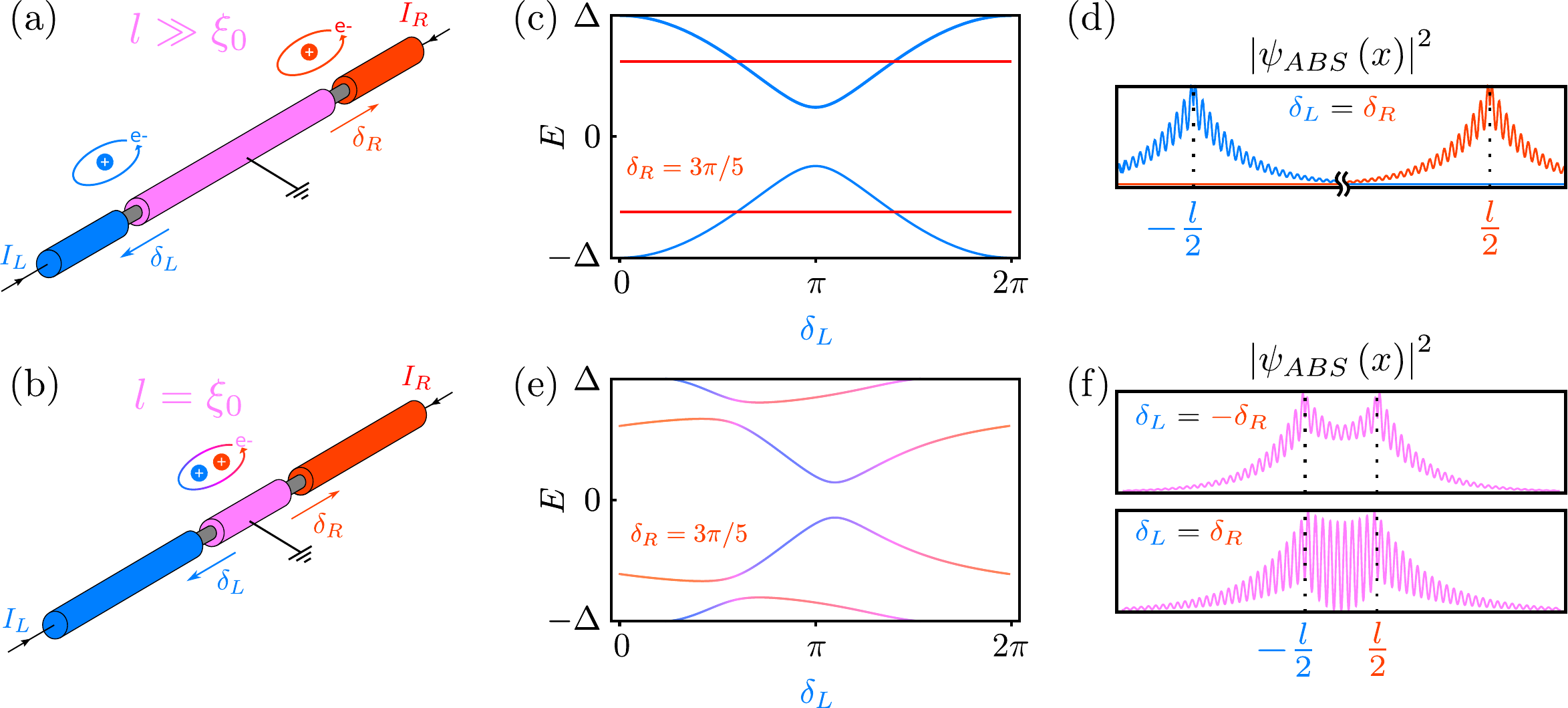}
  \caption{
    Two single-channel short weak links (gray) of transmissions $\tau_L = \tau_R \approx 0.94$ are connected to three superconducting regions (colored), of which the central one (pink) has length $l = 10\xi_0$ (a), or in the case of the Andreev molecule, $l = \xi_0$ (b).
    The ground connection allows applying phase differences $\delta_L$ and $\delta_R$ independently.
    (c) The Andreev Bound State (ABS) spectrum as a function of $\delta_L$ for fixed $\delta_R = 3\pi/5$ for $l = 10\xi_0$ and the corresponding wavefunctions at the two weak links (d).
    The degeneracies at $\delta_L = \pm \delta_R$ are lifted in the spectrum of the Andreev molecule (e) and the wavefunctions hybridize (f).
    Since the phases are $2\pi$ periodic, the $\delta_L = -\delta_R$ degeneracy is shown at $\delta_L = 2\pi-\delta_R = 7\pi/5$.
    Colors indicate relative localization of the wavefunction on the weak links (left: blue, right: red).
}
\label{Fig1}
\end{figure}


The simplest Josephson junction is composed of two superconductors connected by a short, one-dimensional quantum conductor with a single electronic channel.
Experimentally, this has been realized with superconducting atomic contacts~\cite{bretheau_exciting_2013,della_rocca_measurement_2007} and in semiconductor-superconductor nanowires~\cite{krogstrup_epitaxy_2015,albrecht_exponential_2016,deng_majorana_2016,goffman_2017,tosi_2019}.
These experimental results are well described by the Bogolubiov-de-Gennes (BdG) formalism and motivates us to use the same formalism to calculate the junction interaction.
We consider a series connection of two ideal short weak links as shown in \autoref{Fig1}(a) for large separation ($l \gg \xi_0$) and in \autoref{Fig1}(b) for an Andreev molecule ($l = \xi_0$).
By connecting a ground electrode to the central conductor one can flow current independently through each junction or apply individual superconducting phase differences $\delta_{L\left(R\right)}$ using magnetic fields.
Experimentally, achieving small separations $l \lesssim \xi_0$ and making an electrical connection to the central superconductor is feasible with microfabrication techniques since the effective superconducting coherence length is approximately 100~nm for a disordered thin film of aluminum, a typical superconductor.

Within the BdG formalism, electrons in such a circuit can be described by a $2\times2$ Hamiltonian $H$ in Nambu space,
\begin{align}
H &= \left(\begin{array}{cc}
H_{0}+H_{\mathit{WL}} & \Delta\left(x\right)\\
\Delta^{*}\left(x\right) & -H_{0}-H_{\mathit{WL}}
        \end{array}\right),\label{eq:Hamiltonian}
  \\
\Delta(x) &= \begin{cases}
\Delta e^{i\delta_{L}} & \text{if }x<-l/2\\
\Delta & \text{if }|x|<l/2\\
\Delta e^{i\delta_{R}} & \text{if }x>l/2
\end{cases},\label{eq:offdiag}
\end{align}
where $H_{0}=\frac{-\hbar^{2}}{2m}\partial_{x}^{2}-\mu$ is the single particle energy ($m$ is the electron mass, $\mu$ the chemical potential) and $H_{\mathit{WL}}=U_{L}\delta\left(x+l/2\right)+U_{R}\delta\left(x-l/2\right)$ models scattering at the weak links at $x=\pm l/2$ with amplitudes $U_{L\left(R\right)}$.
For simplicity we use Dirac $\delta$-functions for the scatterers, appropriate for weak links of length shorter than $\xi_0$, but expect the effects to persist in longer weak links, diminishing on a length scale given by $\xi_0$.
In the following, we consider symmetric scattering $U_{L} = U_{R} = U_0$ corresponding to both single-channel junctions having transmission $\tau \approx 0.94$, comparable to experimental values~\cite{goffman_2017,tosi_2019}.

The off-diagonal terms of the Hamiltonian, $\Delta(x)$ (\autoref{eq:Hamiltonian}), describe electron pairing in each superconductor.
The superconducting gap amplitude $\Delta$ is constant along the whole device and by gauge invariance we choose the phase of the central superconductor to be zero.
In the one- to few-channel limit considered here the weak link supercurrents are much smaller than the critical currents of the superconducting regions and it is justified to ignore additional spatial variations in the order-parameter amplitude or phase.


Diagonalization of the BdG equation $H\psi=E\psi$ using an evanescent plane-wave basis for the wavefunctions $\psi$ gives a discrete spectrum of Andreev Bound States with energies smaller than the superconducting gap $|E|<\Delta$ and a continuum of states for $|E|>\Delta$.
For the plane-waves we make the standard approximation $\xi_0 \gg \lambda_F$ appropriate for typical superconductors, where $\lambda_F$ is the Fermi wavelength.
Analytic expressions for the spectrum can be obtained at arbitrary separation for opaque junctions ($\tau \ll 1$) as well as for arbitrary transmission at large ($l/\xi_0 \gg 1$) or small separation ($l/\xi_0 \ll 1$).
However the results shown here in the experimentally accessible intermediate case are calculated numerically.

\autoref{Fig1}(c) shows the ABS spectrum for large separation, $l \gg \xi_0$, as a function of the left weak link's phase $\delta_L$ for fixed $\delta_R = 3\pi/5$ across the right weak link.
As expected there is no interaction and the four levels are determined by the single junction expression $E^{\pm}_{A}(\delta,\tau_{0})=\pm\Delta\sqrt{1-\tau_{0}\sin^{2}\frac{\delta}{2}}$ by substituting $\delta = \delta_{L(R)}$ and $\tau_0 = \tau$.
As a consequence of setting $\delta_R = 3\pi/5$, there are two dispersionless ABS (red lines) at energies $E^{\pm}_{A}(3\pi/5,\tau)$.
Since the transmissions of both junctions are equal, $\tau_L = \tau_R = \tau$, the energy levels cross at $\delta_L = \pm \delta_R = \pm 3\pi/5$.
The degeneracy at $\delta_L = -\delta_R$ is actually plotted at $\delta_L = 2\pi-\delta_R = 7\pi/5$ because the phases are $2\pi$ periodic.
We confirm that the spectra correspond to distinct ABS localized at each weak link by plotting the ground-state wavefunctions in~\autoref{Fig1}(d) at the degeneracy point $\delta_L = \delta_R$.

The spectrum of the Andreev molecule~\autoref{Fig1}(e) changes drastically as a result of interactions between the junctions at small separation $l = \xi_0$.
Not only do gaps open and lift the preceding degeneracies, but all spectral lines now depend on $\delta_L$, despite the fact that $\delta_R$ remains fixed.
Additional spectra for separation $l = 2\xi_0, l = 0.5\xi_0$, and $l \approx 0$ can be found in the Supporting Information.

The breaking of symmetry about $\delta_L = \pi$, absent in the spectra of the isolated junctions, is permitted by time-reversal invariance, which only requires $E(\delta_{L},\delta_{R})=E(-\delta_{L},-\delta_{R})$
and not $E(\delta_{L},\delta_{R})=E(-\delta_{L},\delta_{R})$.
Although the continuum states are not plotted, states near the gap edge also develop a dispersion in the regions where the bound states approach the gap.

The opening of gaps is reminiscent of a coupling which hybridizes previously orthogonal states, and this picture is confirmed by a plot of the ABS wavefunctions in~\autoref{Fig1}(f) for $\delta_L = \pm\delta_R$.
Only the wavefunction of the negative-energy state closest to zero energy is plotted in each case.
The rapid oscillations here and in~\autoref{Fig1}(d) are due to the propagating part of the plane waves, whose wavelength  $\lambda_F = \xi_0/5$ is chosen for visibility.
The ABS wavefunctions are spread over both junctions with a significant weight in the central superconductor.
Due to the off-diagonal term $\Delta(x)$ (\autoref{eq:offdiag}) the parity symmetry of the Hamiltonian is not the same for $\delta_L = \delta_R$ and $\delta_L = -\delta_R$.
This results in the asymmetric spectrum as well as the differences in symmetry between the two wavefunctions, similar to bonding and anti-bonding states of a molecule.

In the context of quartets, the distinct hybridization mechanisms at $\delta_L = \delta_R$ and $\delta_L = -\delta_R$ are called double crossed Andreev reflection (dCAR) and double elastic cotunneling (dEC), respectively~\cite{freyn_production_2011,feinberg_quartets_2015}.
Microscopically, dCAR corresponds to the transfer of Cooper pairs from the inner superconducting electrode to the outer ones while dEC corresponds to the transfer of Cooper pairs directly from the left to right electrode.
One can link this Cooper pair transfer picture to our formalism via the scattering matrix approach as well as obtain analytical expressions for the dCAR and dEC currents in special limiting cases (see Supporting Information), but it is not necessary to invoke entanglement as with the quartet interpretation.

\begin{figure}
\includegraphics[width=0.8\columnwidth]{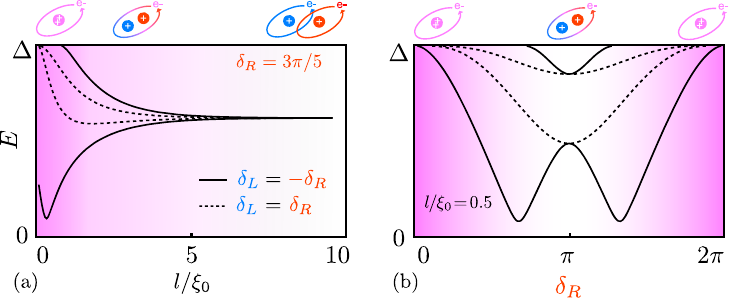}
\caption{
  (a) Positive Andreev Bound State spectrum at symmetry points $\delta_{L}=-\delta_{R}$ (solid lines) and $\delta_{L}=\delta_{R}$ (dashed lines) as a function of $l/\xi_{0}$ for fixed $\delta_{R}=3\pi/5$.
  Due to hybridization the energy levels split as the separation is reduced and some states are pushed into the continuum.
  (b) The level splitting, as well as the number of bound states, can also be controlled by adjusting $\delta_{R}$ for fixed separation, $l/\xi_0 = 0.5$.
  Oscillations on a scale $\lambda_F$ are avoided by choosing $k_{F}l=\pi/2$ (mod $2\pi$) with $k_{F}l\gg1$.
}
\label{Fig2}
\end{figure}

The transition from two independent junctions ($l \gg \xi_0$) to a single one ($l = 0$) with an intermediate molecular state is visualized in \autoref{Fig2}(a) where the positive spectral lines for $\delta_{L}=\pm\delta_{R}$ are plotted as a function of $l/\xi_{0}$ for $\delta_{R}=3\pi/5$ and the same transmission as before.
At large separation $l/\xi_{0}\gg1$ all positive ABS converge to the same energy, $E^{+}_{A}(\delta=3\pi/5,\tau)$, that of the upper red line in~\autoref{Fig1}(c).

For the hybridization mechanism at $\delta_{L}=-\delta_{R}$ (dEC, solid lines), with decreasing $l/\xi_{0}$ the two degenerate ABS of the left and right junction gradually split to form bonding and anti-bonding molecular states until the higher-energy ABS escapes into the continuum.
The ABS pushed into the continuum is referred to as a ``leaky'' Andreev state \cite{bagwell_suppression_1992} but unlike in-gap ABS are delocalized plane waves~\cite{olivares_dynamics_2014}.
Beyond that point the device hosts only one pair of ABS, symmetric in energy, such that it behaves as a single artificial atom rather than a molecule.
Incidentally, around this transition from molecule to atom, the lower ABS reaches a minimum resulting in an overall shape which suggests the interatomic potentials used to describe the formation of molecules.
Physically, this minimum occurs due to competition between hybridization which pushes the lower ABS toward zero energy and scattering which pushes it away (see Supporting Information).

At $l/\xi_{0}=0$, in contrast to a real interatomic potential, the energy of this lower state does not diverge and is less than the energy at large separation, $l/\xi_{0}\gg1$.
The full spectrum for $l/\xi_{0}=0$ can be found in the Supporting Information. 

For the dashed $\delta_{L}=\delta_{R}$ (dCAR) spectral lines, the level splitting is weaker than for dEC (solid lines) and both states are pushed into the continuum as $l/\xi_{0}$ approaches zero.
As with the level splitting, the depth of the potential well for the lower dashed line is smaller than for dEC (lower solid line).
This is because the dCAR process is in general weaker than dEC for high transmissions.

A more experimentally feasible possibility to tune the hybridization strength is adjusting $\delta_{R}$ at fixed separation, as shown in \autoref{Fig2}(b) for $l/\xi_{0}=0.5$.
Different regimes can be identified by the number of bound states.
As $\delta_{R}$ increases from $0$ to $\pi$, states gradually detach from the continuum and form bound states.
The size of the level splitting of the $\delta_L = \delta_R$ line (dCAR, dashed) is maximal at $\delta_{R}=\pi$ whereas for $\delta_L = -\delta_R$ (dEC, solid line) this maximum occurs at an intermediate value.
By changing $\delta_{R}$ one can control the degree of hybridization and tune the molecular nature of the system.

As a function of the transmission $\tau$, the general tendency is for the energy splitting at $\delta_L = -\delta_R$ to monotonically increase with $\tau$ whereas the splitting at $\delta_L = \delta_R$ has a peak and goes to zero at $\tau = 1$.

The Andreev molecule, as a tunable multi-level qubit, may be useful for the preparation and manipulation of quantum states.
In an Andreev molecule with moderate transmission and separation, one could envision a protocol where the system, initially in the ground state, is tuned via the phases $\delta_{L,R}$ to a level splitting in the gigahertz range.
Microwave pulses could then prepare an excited state which would be protected by subsequently detuning the phases away from the splitting where transitions are less likely.
Although coherence times of Andreev-based qubits~\cite{tosi_2019,hays_2018} are currently an order of magnitude lower than that of conventional superconducting qubits, the decoherence mechanisms are not understood and may well be mitigated in the future.



The remarkable features of the spectrum of the Andreev molecule, in particular the asymmetry and additional energy dispersion in~\autoref{Fig1}(e), dramatically modifies the Josephson effect and gives rise to non-local Josephson supercurrents.
These supercurrents are calculated from the eigenstates of the BdG equation~\cite{blonder_transition_1982,beenakker_josephson_1991,bardeen_1969} in a manner analogous to calculating probability currents.
Both Andreev bound states and continuum states contribute significantly to the total supercurrent, unlike the case for isolated junctions where the continuum plays no role~\cite{beenakker_josephson_1991}.

\begin{figure}
\includegraphics[width=0.8\columnwidth]{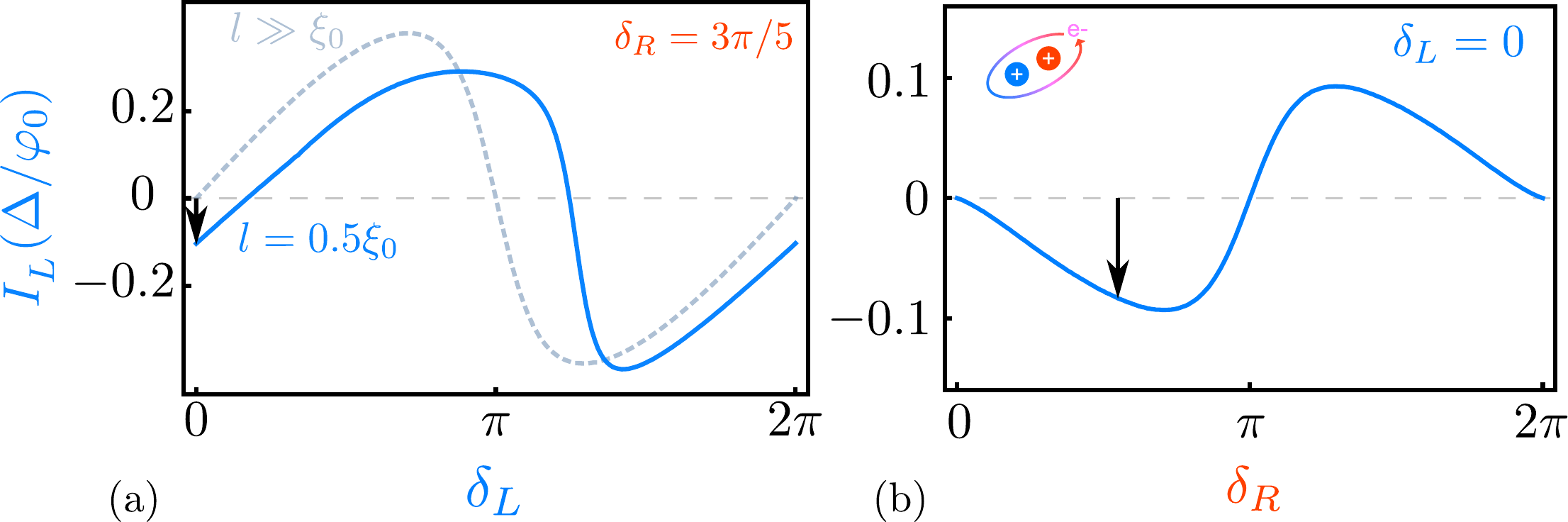}
\caption{Andreev molecule current-phase relation and $\varphi$-junction behavior.
  (a) The current-phase relation of the left weak link, $I_{L}(\delta_{L},\delta_{R}=3\pi/5)$, is shifted and distorted for small separation ($l = 0.5\xi_0$, solid line) as compared to the isolated junction ($l \gg \xi_0$, dashed line).
  (b) The zero-phase supercurrent $I_L(\delta_L = 0)$ (black vertical arrow) is plotted as a function of $\delta_R$, showing a tunable $\varphi$-junction effect.
  Currents are normalized to $\Delta/\varphi_0$, with $\varphi_0 = \hbar/2e$ the reduced flux quantum.
}
\label{Fig3}
\end{figure}

\autoref{Fig3}(a) shows the supercurrent-phase relation of the left junction $I_{L}\left(\delta_{L},\delta_{R}\right)$ for a fixed superconducting phase difference $\delta_{R}=3\pi/5$ across the right junction.
Plots are given for an isolated junction (dashed line) and for an Andreev molecule with $l = 0.5\xi_0$ (solid line).
The curve for small separation is shifted with respect to that for large separation, and increases as the separation $l/\xi_0$ is reduced.
In addition there is a distortion to the curve which breaks the odd symmetry in $\delta_L$ about the zeroes in the current.
Furthermore, the positive and negative critical currents are not equal, and although not shown here, have magnitudes which depend on $\delta_R$. 
These modifications, resulting from the dependence of $I_L$ on $\delta_R$ for $l \lesssim \xi_0$, are non-local effects in that the phase drop $\delta_R$ occurs away from the left weak link, at a distance on the order of $\xi_0$, which can be hundreds of nanometers depending on the superconductor.

For an isolated, time-reversal invariant Josephson junction,  \autoref{Fig1}(a), the supercurrent must be zero at a phase difference of $0$ and $\pi$, \autoref{Fig3}(a) (dashed line)~\cite{golubov_2004}.
But for the Andreev molecule where $I_L$ depends on $\delta_R$, since time-reversal symmetry only imposes $I_{L}\left(-\delta_{L},-\delta_{R}\right)=-I_{L}\left(\delta_{L},\delta_{R}\right)$, the current $I_{L}$ can be non-zero at $\delta_{L}=0$.
A non-zero supercurrent at zero phase difference is the hallmark of a $\varphi$-junction, in which the current crosses zero at a phase $\varphi$ instead of zero~\cite{koshelev_2003}.

In~\autoref{Fig3}(b) the zero-phase supercurrent $I_L^0 = I_L(\delta_L=0,\delta_R)$ is plotted as a function of $\delta_R$.
The sign of $I_{L}^0$ for small $\delta_R$ is negative, indicating that the current flows in the same direction as the phase difference $\delta_{R}$, from left to central electrode in~\autoref{Fig1}(b).
This is akin to a ``ferromagnetic'' interaction in that the current $I_{L}^{0}$ and phase difference $\delta_R$ are aligned.
This tendency is explained by the fact that for vanishing separation $l$, there is only one junction with an overall phase drop given by $\delta_L-\delta_R$ and therefore $I_L(\delta_L,\delta_R) = I_L(\delta_L-\delta_R)$.
In other words, the $\varphi$-junction phase offset approaches $\varphi=-\delta_R$ in the limit $l/\xi_0 \rightarrow 0$.




\begin{figure}
\includegraphics[width=0.8\textwidth]{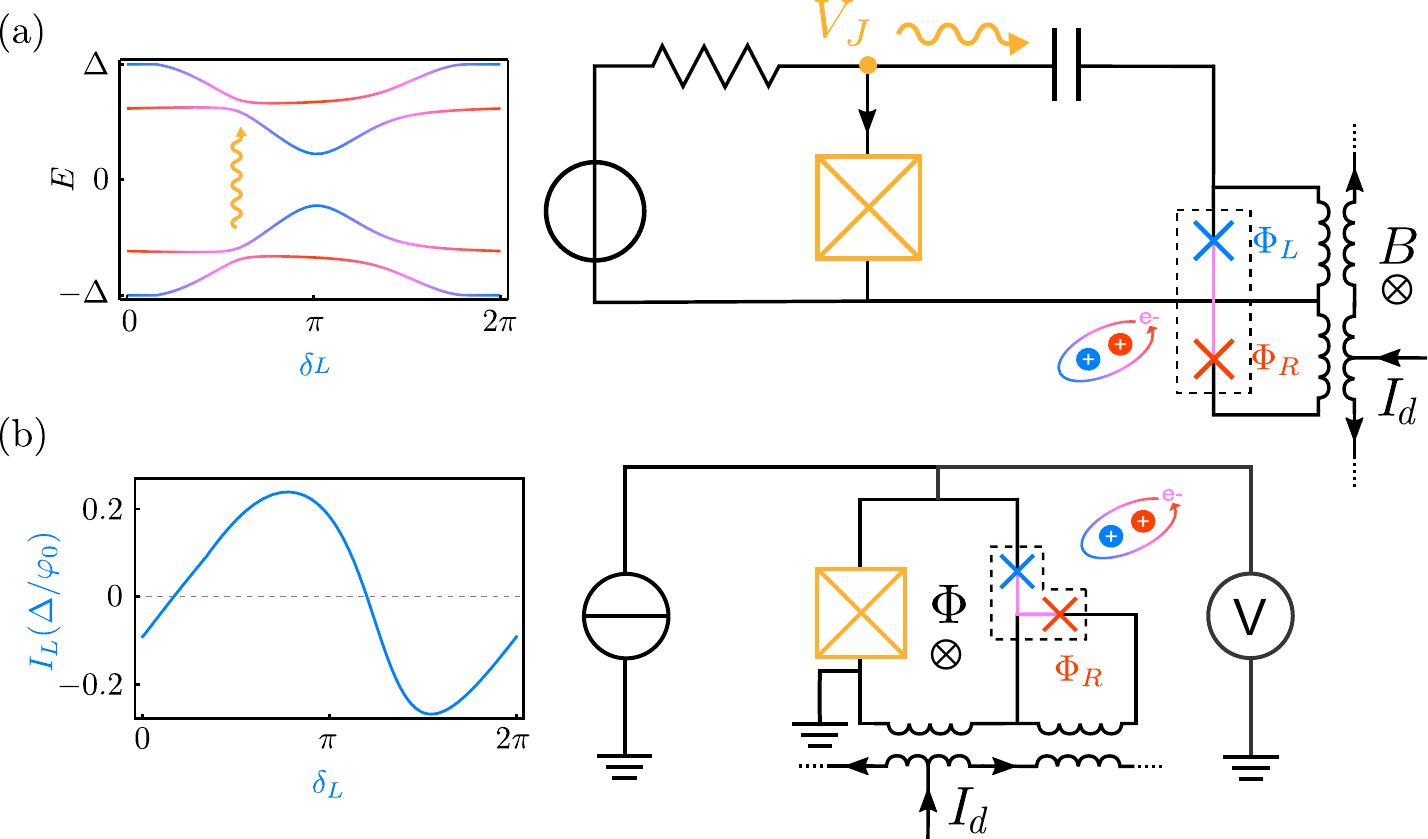} 
\caption{Experimental proposals. (a)
    The Josephson tunnel junction biased at $V_J$ (boxed cross) acts as a spectrometer for transitions of the Andreev molecule (colored crosses).
    (b) Setup for switching-current measurements of the current-phase relation of an Andreev molecule (described in text).
    An overall magnetic field $B$ and control current $I_d$ allow independent control of $\delta_{L(R)}$ via the loop fluxes $\Phi,\Phi_{L(R)}$.
}
\label{Fig4}
\end{figure}

The two main signatures of Andreev molecules at equilibrium are the spectral avoided crossings (\autoref{Fig2}) and the non-local current-phase relation (\autoref{Fig3}).
For large effects, the Andreev molecule should have small separation ($l \lesssim \xi_0$) and high transmission ($\tau \lesssim 1$), realistic conditions for epitaxial semiconductor-superconductor devices~\cite{goffman_2017,tosi_2019}.
Additional calculations have shown that these effects are robust to disorder and persist if there are multiple conduction channels.
Andreev molecules can be detected with experiments at zero voltage bias, as described below.
In non-equilibrium dc transport experiments, where junctions are voltage biased~\cite{freyn_production_2011,pfeffer_subgap_2014,cohen_nonlocal_2018}, these signatures are not present and dynamic processes such as mutual phase locking~\cite{likharev_dynamics_1986} complicate interpretation.

In~\autoref{Fig4}(a), we propose a setup to perform spectroscopy of an Andreev molecule in a similar manner to the spectroscopy of an Andreev ``atom'' in a superconducting atomic contact \cite{bretheau_exciting_2013}.
A large tunnel Josephson junction (yellow) is simultaneously used as an on-chip microwave source and detector.
When a voltage bias $V_{J}$ is applied across this spectroscopy junction, it emits microwave current at frequency $\nu=2eV_{J}/h$ via the AC Josephson effect.
This high-frequency current is coupled via the capacitor to the Andreev molecule, and due to inelastic Cooper-pair tunneling, will give peaks in the DC current-voltage characteristic of the spectrometer junction, $I_{J}(V_{J})$, at voltages $h\nu/2e = E_{T}/2e$ corresponding to the transitions in the Andreev molecule spectrum, \autoref{Fig1}(e), of energy $E_{T}$.
The magnetic field $B$ and gradiometric control current $I_{d}$ allow independently tuning the phases $\delta_{L(R)}$ via the magnetic fluxes $\Phi_{L(R)}$, assuming that loop inductances are negligible compared to the Josephson inductance of the weak links.
A full measurement $I_{J}(V_{J},B,I_{d})$ would allow reconstructing the spectra of~\autoref{Fig1}(e) as well as verifying the predictions of~\autoref{Fig2}(b).

Josephson spectroscopy is well suited for detecting Andreev molecules, as $E_{T}$ may be comparable to $\Delta/h$ (90 GHz for aluminum), well within the spectrometer bandwidth. Conventional microwave spectroscopy~\cite{janvier_coherent_2015,tosi_2019}, with greater sensitivity but narrower range, may also be used, as well as tunneling spectroscopy~\cite{pillet_andreev_2010,deng_2018} instead of photonic spectroscopy.

In a second experiment, the existence of a non-local current-phase relation can be directly determined by measurements of the switching current.
This type of experiment has been performed on superconducting atomic contacts~\cite{goffman_supercurrent_2000}, graphene~\cite{english_observation_2016} and carbon nanotubes~\cite{delagrange_0-_2016}.
As shown in~\autoref{Fig4}(b), a large Josephson junction of critical current $I_{0}\gg\textrm{\textrm{max}}\left(I_{L},I_{R}\right)$ (boxed cross) is wired in parallel with the left weak link, $L$ (upper black cross), of the Andreev molecule, forming an asymmetric SQUID.
Similarly to the previous proposal, $\delta_L$ and $\delta_R$ are controlled independently using the flux $\Phi$ and control current $I_d$. 
Due to the asymmetry, the SQUID critical current $I_{SQ}$ is essentially given by that of the large junction, effectively at a phase difference of $\pi/2$, modulated by the supercurrent of the $L$ junction, $I_{SQ}\approx I_{0}+I_{L}(\delta_L,\delta_{R})$, where flux quantization constrains $\delta_{L}=\Phi/\varphi_0-\pi/2$.
By sending current pulses or ramps and measuring the switching of the SQUID to a non-zero voltage state, one can extract the current-phase relation $I_{L}(\delta_L,\delta_R)$ and demonstrate that there is a non-local component that depends on $\delta_{R}$.










We have shown that two superconducting weak links separated by a distance on the order of the superconducting coherence length exhibit a non-local Josephson effect, a phenomenon attributed to the hybridization of Andreev Bound States between the two weak links and the formation of an Andreev molecule.
Our work provides a microscopic understanding of the ``order-parameter,'' or Cooper-pair mediated, interaction between Josephson junctions.
The most prominent signatures of non-local effects are a modification of the Andreev Bound State energy spectrum and the emergence of a $\varphi$-junction, a weak link with non-zero supercurrent at zero phase difference.
Such signatures can be detected with well-established spectroscopy techniques and switching-current measurements.

The Andreev molecule, with tunable energy gaps, may serve as a new type of superconducting qubit, and its $\varphi$-junction behavior may find use in magnetometry.
In addition the large junction interaction holds promise for applications employing arrays of coupled weak links, such as in quantum information, metrology, or sensing.
Arrays of junctions in the Andreev molecule limit ($l \lesssim \xi_0$) may achieve new limits of qubit coupling, quantum amplification, voltage stability, or magnetic field sensitivity.
Such arrays may also serve as quantum simulators of polymers or to implement model Hamiltonians.
By controlling the separation between weak links in the array as well as the individual phase differences, the strength of the coupling between Andreev states at adjacent weak links can be varied.
For example, alternating short and long separation would allow generating a Su-Schrieffer-Heeger (SSH) type Hamiltonian modeling a polyacetylene molecule.
It may also be possible to introduce repulsive on-site interactions by using chains of weak links where charging energy is important.

The predicted coupling between two weak links should also exist for exotic Andreev Bound States where spin-orbit and Zeeman effects play a role, and could be exploited for topological quantum computation with Majorana bound states.



\begin{acknowledgement}

  We acknowledge fruitful discussions with Yuli Nazarov, Régis Mélin, Benoît Douçot, Yonatan Cohen, Yuval Ronen, Hugues Pothier, Fabien Lafont, Landry Bretheau and Benjamín Alemán and support from Jeunes Equipes de l'Institut de Physique du Collège de France.
  This research was supported by IDEX grant ANR-10-IDEX-0001-02 PSL and a Paris ``Programme Emergence(s)'' Grant.
  This project has received funding from the European Research Council (ERC) under the European Union's Horizon 2020 research and innovation programme (grant agreement 636744).

\end{acknowledgement}

\begin{suppinfo}
Additional discussion on separation dependence of spectra; supplementary spectra of Andreev molecules for different values of separation $l$; details of the calculation of the Andreev molecule spectrum and supercurrent within the Bogolubiov-de-Gennes formalism; discussion of Andreev molecules with Josephson tunnel junctions.
\end{suppinfo}
\bibliography{MyLibrary.bib}

\end{document}